\documentclass[conference,letterpaper]{IEEEtran}

\usepackage[utf8]{inputenc} 
\usepackage[T1]{fontenc}
\usepackage{url}
\usepackage{ifthen}
\usepackage{cite}
\usepackage{amsthm}
\usepackage{graphicx}
\usepackage[cmex10]{amsmath}

\usepackage{xspace}
\usepackage{enumitem} %
\usepackage{etoolbox}
\usepackage{csquotes} %
\usepackage{verbatim}
\usepackage{mathtools}

\usepackage{graphicx}
\usepackage[dvipsnames]{xcolor}
\usepackage{xcolor}
\usepackage{tikz} 
\usepackage{tikz-cd} 

\usepackage{amssymb}
\usepackage{bbm}   
\usepackage{mathrsfs}
\usepackage{mathtools} %

\usepackage{booktabs} %
\usepackage{diagbox}  %
\usepackage{multirow} %
\usepackage{wrapfig} %
\usepackage{longtable} %

\usepackage{cancel}

\allowdisplaybreaks
\usetikzlibrary{arrows}
\usetikzlibrary{trees}

\definecolor{CommentBlue}{RGB}{0,80,239}
\definecolor{CommentPurple}{RGB}{145, 32, 186}
\definecolor{CommentRed}{RGB}{168,24,24}
\definecolor{CommentGreen}{RGB}{15,100,15}
\definecolor{CommentOrange}{RGB}{204,102,0}
\definecolor{CommentAltRed}{RGB}{200,12,12}
\definecolor{lgray}{gray}{0.9}

\DeclareMathAlphabet\mathbfcal{OMS}{cmsy}{b}{n}
\def\mc{\mathcal}

\def\mbf{\mathbf}

\newcommand{\bfs}{\mathbf{s}}

\newcommand{\bfx}{\mathbf{x}}
\newcommand{\bfy}{\mathbf{y}}

\newcommand{\cA}{{\mathcal A}}

\newcommand{\cO}{{\mathcal O}}

\newcommand{\cS}{{\mathcal S}}

\newcommand{\cX}{{\mathcal X}}
\newcommand{\cY}{{\mathcal Y}}
\newcommand{\cZ}{{\mathcal Z}}

\newcommand{\vs}{\underline{s}}

\newcommand{\vx}{\underline{x}}
\newcommand{\vy}{\underline{y}}
\newcommand{\vz}{\underline{z}}

\DeclarePairedDelimiter\braces{\lbrace}{\rbrace}
\DeclarePairedDelimiter\bracks{\lbrack}{\rbrack}

\newtheorem{remark}{Remark}
\newtheorem{definition}{Definition}
\newtheorem{theorem}{Theorem}

\newcommand{\N}{\mathbb{N}}

\newcommand{\defeq}{\triangleq}
\newcommand{\coloneqq}{\triangleq}

\newcommand{\Indic}[1]{\mathbbm{1}\braces*{#1}} %

\renewcommand{\vec}[1]{\underline{#1}}

\DeclareMathOperator{\poly}{poly}

\newcommand{\simplex}{\Delta}

\NewDocumentCommand{\expect}{ e{^} s o >{\SplitArgument{1}{|}}m }{%
  \operatorname{\mathbb{E}}%
  \IfValueT{#1}{{\!}^{#1}}%
  \IfBooleanTF{#2}{%
    \expectarg*{\expectvar#4}%
  }{%
    \IfNoValueTF{#3}{%
      \expectarg{\expectvar#4}%
    }{%
      \expectarg[#3]{\expectvar#4}%
    }%
  }%
}
\NewDocumentCommand{\expectvar}{mm}{%
  #1\IfValueT{#2}{\nonscript\;\delimsize\vert\nonscript\;#2}%
}
\DeclarePairedDelimiterX{\expectarg}[1]{[}{]}{#1}
\newcommand{\E}{\ensuremath{\mathbb{E}}}
\NewDocumentCommand{\prob}{ e{^} s o >{\SplitArgument{1}{|}}m }{%
  \operatorname{\mathbb{P}}%
  \IfValueT{#1}{{\!}^{#1}}%
  \IfBooleanTF{#2}{%
    \probarg*{\probvar#4}%
  }{%
    \IfNoValueTF{#3}{%
      \probarg{\probvar#4}%
    }{%
      \probarg[#3]{\probvar#4}%
    }%
  }%
}
\NewDocumentCommand{\probvar}{mm}{%
  #1\IfValueT{#2}{\nonscript\;\delimsize\vert\nonscript\;#2}%
}
\DeclarePairedDelimiterX{\probarg}[1]{(}{)}{#1}

\NewDocumentCommand{\var}{ e{^} s o >{\SplitArgument{1}{|}}m }{%
  \operatorname{\mathrm{Var}}
  \IfValueT{#1}{{\!}^{#1}}
  \IfBooleanTF{#2}{%
    \vararg*{\varvar#4}%
  }{%
    \IfNoValueTF{#3}{%
      \vararg{\varvar#4}%
    }{%
      \vararg[#3]{\varvar#4}%
    }%
  }%
}
\NewDocumentCommand{\varvar}{mm}{%
  #1\IfValueT{#2}{\nonscript\;\delimsize\vert\nonscript\;#2}%
}
\DeclarePairedDelimiterX{\vararg}[1]{[}{]}{#1}

\newcommand{\condon}{\,\ifnum\currentgrouptype=16 \middle\fi|\,} %

\DeclareMathOperator{\supp}{supp}

\newcommand{\type}[1]{T_{#1}}		%
\newcommand{\typeclass}[2]{\mathcal{T}^{#1}(#2)}	%

\newcommand{\rs}{\mbf{s}} %
\newcommand{\rx}{\mbf{x}} %
\newcommand{\ry}{\mbf{y}} %
\newcommand{\rvx}{\vec{\mbf{x}}}
\newcommand{\rvy}{\vec{\mbf{y}}}
\newcommand{\rvs}{\vec{\mbf{s}}}

\newcommand{\xtypes}{\Gamma}

\newcommand{\stypes}{\Lambda}

\newcommand{\msgerr}{\varepsilon}
\newcommand{\maxerr}{\hat{\varepsilon}}
\newcommand{\avgerr}{\bar{\varepsilon}}

\newcommand{\jamset}{\mathfrak{J}}

\newcommand{\jam}{J_{\rvs}}

\newcommand{\dec}{\psi}
\newcommand{\renc}{\boldsymbol{\phi}}

\newcommand{\removed}[1]{}

\usepackage{mleftright}

\renewcommand{\brack}[1]{\mleft[#1\mright]}

\newcommand{\CN}{\mathrm{ECN}}

\newcommand{\bigO}{\cO}

\newcommand{\xwin}{w_x}
\newcommand{\swin}{w_s}
\newcommand{\winratio}{\alpha}
\newcommand{\AVC}{\cA}
\newcommand{\winAVC}{\AVC_{\textrm{win}}}
\newcommand{\Capobl}{C_{\textrm{obl}}}
\newcommand{\CapAobl}{\bar{C}_{\textrm{obl}}}
\newcommand{\CapMobl}{\hat{C}_{\textrm{obl}}}
\newcommand{\Capran}{C_{\textrm{ran}}}
\newcommand{\Caplist}{C_{\textrm{list}}}
\newcommand{\CaplistA}{\bar{C}_{\textrm{list}}}
\newcommand{\CaplistM}{\hat{C}_{\textrm{list}}}
\newcommand{\win}{{\textrm{win}}}
\newcommand{\CapwinA}{\bar{C}_\win}
\newcommand{\CapwinM}{\hat{C}_\win}
\newcommand{\Capwin}{C_\win}

\newcommand{\runstep}{l}

\usepackage[color=red!30,colorinlistoftodos]{todonotes}

\newcommand{\bikd}[1]{\textcolor{CommentGreen}{{\bf Bikash}: {\em #1}}}

\begin{document}
\title{Sliding Window Adversarial Channels}

\author{%
  \IEEEauthorblockN{B. K. Dey}
  \IEEEauthorblockA{%
                    bikash@ee.iitb.ac.in}
  \and
  \IEEEauthorblockN{S. Jaggi}
  \IEEEauthorblockA{%
                    sid.jaggi@bristol.ac.uk}
  \and
  \IEEEauthorblockN{M. Langberg}
  \IEEEauthorblockA{%
                    mikel@buffalo.edu}
  \and
  \IEEEauthorblockN{A. D. Sarwate}
  \IEEEauthorblockA{%
                    ads221@rutgers.edu}
  \and
  \IEEEauthorblockN{Y. Zhang}
  \IEEEauthorblockA{%
                    zephyr.z798@gmail.com}
}

\maketitle

\begin{abstract}
In an arbitrarily varying channel (AVC), the channel has a state which is under the control of an adversarial jammer and the corresponding capacities are often functions of the ``power'' constraints on the transmitter and jammer. In this paper we propose a model in which the constraints must hold almost surely over contiguous subsequences of the codeword and state, which we call a sliding window constraint. We study oblivious jammers and codes with stochastic encoding under maximum probability of error. 
We show that this extra limitation on the jammer is beneficial for the transmitter: in some cases, the capacity for unique decoding with a sliding window constraint is equal to the capacity for list decoding in the standard model without sliding windows, 
roughly implying that the addition of window constraints reduces list decoding to unique decoding.
The list decoding capacity in the standard model can be strictly larger than the
unique decoding capacity.
\end{abstract}

\section{Introduction}

In a discrete adversarial channel model, a transmitter (Alice) encodes a message $m$ into a codeword $\vx$ and sends it over a channel $W_{\bfy|\bfx,\bfs}(y|x,s)$ to a receiver/decoder (Bob). The channel ``noise'' (state) $\vs$ is partially controlled by a jammer (James), whose objective is to cause Bob to make a decoding error. There are several varieties of such models which can be formulated as variations on arbitrarily varying channels (AVCs)~\cite{DeyJLSZ:24monograph}. Codes and capacity results become more nuanced in the cases where Alice's encoding or James' jamming strategy must satisfy (cost) constraints: for blocklength $n$ codes, the type of $\vx$ or $\vs$ is constrained to lie in some pre-specified set, often defined in terms of a per-letter cost.

In this paper we revisit the cost constraint formulation and consider \emph{sliding window constraints} in which the type of any contiguous window of codeword entries and/or any contiguous window of state variables must lie in pre-specified sets. The window lengths need not be the same for Alice and James: for blocklength $n$, $\vx$ may be constrained by windows of length $\xwin(n)$ and $\vs$ by windows of length $\swin(n)$. This formulation captures a natural restriction that the total energy expended by Alice or James over a certain length of time should not exceed a threshold. We provide results for the situation where windows are neither too long nor too short, namely $\xwin(n),\swin(n) \in [\omega(\log n), o(n)]$. 

The basics of AVCs are summarized in a recent monograph~\cite{DeyJLSZ:24monograph}. We study oblivious adversaries without common randomness but allow Alice to use private randomness, unknown to Bob. %
In the standard windowless model, Csisz\'{a}r and Narayan~\cite{CN:88constraints:deterministic} proved a single-letter formula for the average-error capacity under average per-letter cost constraints on Alice and James's inputs that hold over the entire blocklength $n$. They showed that the capacity is a max-min of a mutual information between the input and output of the channel. A key idea in their converse is that the capacity is 0 if James can ``symmetrize'' the channel (a notion originally due to~Ericson~\cite{Ericson:85exponent}): this means that Alice's choice of input distribution must satisfy both her own cost constraint and a condition of ``non-symmetrizability.'' 
This implies that the capacity for deterministic codes may be positive and smaller than the capacity for randomized codes, which differs from the unconstrained case~\cite{ahlswede-1978}.

Our Theorem \ref{thm:eq} shows that if a windowless AVC $\AVC$ is non-symmetrizable (Definition \ref{def:CN-sym}) %
then the capacity $\Capwin(\winAVC)$ of the windowed AVC $\winAVC$ equals the \emph{list decoding capacity} $\Caplist(\AVC)$ for $\AVC$ with list size $\poly(n)$. The list decoding capacity can be strictly larger than the unique decoding capacity.  If James has a tighter window constraint than Alice ($\xwin(n) > \swin(n)$), our Theorem \ref{thm:wxgtws} 
presents improved sufficient conditions for which $\Capwin(\winAVC)=\Caplist(\AVC)$

Our model is similar to the Masters thesis by Liu~\cite[Section 2.5, Chapter 3]{Liu:87thesis}, which %
studied randomized coding for Gaussian AVCs and discrete AVCs under a limited form of window constraints.
In the non-oblivious setting for channels over large alphabets, inner and outer bounds for sliding window erasures were obtained by Leong and Ho~\cite{Leong_erasure} and Tekin et al~\cite{Tekin_erasure}, and for sliding window errors in Gelles et al.~\cite{Gelles_erasure}.

\section{Channel Model}

For $n \in \mathbb{N}$, $[n] = \{1,2,\ldots, n\}$. For a finite set $\cZ$ and a vector $\vz  = (z_1, z_2, \ldots, z_n) \in \cZ^n$, the subvector $\vz_{[a:b]} = (z_a,z_{a+1},\ldots,z_{b-1})$. 
The type of $\vz$ is $\type{\vz}$ and the set of blocklength-$n$ types on $\cZ$ is $\typeclass{n}{\cZ}$. The set of probability mass functions on $\cZ$ is $\simplex(\cZ)$. For a joint distribution $U_{\bfs|\bfx}P_\bfx \in \simplex(\cS \times \cX)$, the notation $\brack{U_{\bfs|\bfx} P_\bfx}_\bfs \in \stypes$ is the marginal distribution on $\bfs$.

Let $\cX$, $\cS$, and $\cY$ be finite discrete sets (alphabets). Let $W_{\ry|\rx,\rs}(y | x, s)$ be a channel with input $x \in \cX$, output $y \in \cY$, and state $s \in \cS$. If $\vx = (x_1, x_2, \ldots, x_n)$, $\vy = (y_1, y_2, \ldots, y_n)$ and $\vs = (s_1, s_2, \ldots, s_n)$ are length $n$ vectors, the probability of observing the output $\vy$ given the input $\vx$ and state $\vs$ over $W$ is given by:
	\begin{align}
	W_{\rvy|\rvx,\rvs}(\vy | \vx, \vs) = \prod_{i=1}^{n} W_{\ry|\rx,\rs}(y_i | x_i, s_i).
	\label{eq:blockchannel}
	\end{align}
We constrain the input and state sequences to lie in convex sets. Let $\xtypes \subset \simplex(\cX)$ and $\stypes \subset \simplex(\cS)$ be two convex sets and $\vx \in \xtypes$ and $\vs \in \stypes$. We define an AVC as the tuple $\AVC = (W_{\ry|\rx,\rs}, \xtypes,\stypes)$. %
In this paper we introduce a variant of an AVC, which we call a \emph{windowed AVC}, in which the codewords and state vectors must satisfy additional constraints. Let $\xwin(n)$ and $\swin(n)$ be two functions from $\N \to \N$ such that the \emph{window ratio function} $\winratio(n) = \frac{\swin(n)}{\xwin(n)}$ has a limit $\winratio^* = \lim_{n \to \infty} \winratio(n)$, which we call the \emph{window aspect}. A windowed AVC $\winAVC$ is then a tuple $(W_{\ry|\rx,\rs}, \xtypes, \stypes, \xwin, \swin)$.

An $(n,M)$ code for a windowed AVC $\winAVC$ is a pair of maps $(\renc,\dec)$, where the encoder $\renc \colon [M] \to \cX^n$ is a (probabilistic) map such that any contiguous window of length $\xwin(n)$ satisfies the cost constraint, i.e., for $\forall m\in [M]$ and $\forall i \in [n-w_x(n)]$,
    $\type{\renc(m)_{[i:(i+\xwin(n))]}} \in \xtypes$,
\noindent and the decoder $\dec$ is a map $\dec \colon \cY^n \to [M]$. 
We denote the decoding region for message $m$ as $\mc{D}_m = \{ \vy : \dec(\vy) = m \}$. The rate of the code is $R = \frac{1}{n} \log_2(M)$ bits per channel use.

We assume James is an \emph{oblivious} adversary~\cite{DeyJLSZ:24monograph} who knows the maps $\renc$ and $\dec$ but not the realization of $\renc$ or the transmitted codeword $\renc(m)$. A jamming strategy $\jam$ for a windowed AVC $\winAVC$ is a distribution on admissible state vectors such that any contiguous window of length $\swin(n)$ in the state vector must satisfy the cost constraint, i.e., $\forall \vs \in \supp(\jam)$ and $\forall i \in [n-w_s(n)]$,
    $\type{\vs_{[i:(i+\swin(n))]}} \in \stypes$. %
Denote by $ \jamset(\stypes) $ the set of all such jamming strategies. 

The inputs $\rvx(m) \defeq \renc(m)$ and $\rvs$ produce the output $\rvy$ according to the distribution in \eqref{eq:blockchannel}. The probability of correct decoding for a fixed $(m,\vx,\vs)$ is
     \begin{align*}
     \nu(m,\vx,\vs) = \sum_{\vy \in \mc{D}_m} W_{\ry | \rx, \rs}(\vy | \vx, \vs) \jam(\vs)  \Indic{ \renc(m) = \vx }.
     \end{align*}
The error probability for message $m \in [M]$ and state sequence $\vs$ is
 	\begin{align}
 	\msgerr(m,(\renc,\dec),\vs) = 1-\E_{\renc} \bracks*{  \sum_{\vx,\vs} \nu(m,\vx,\vs) }.            
 		\label{eq:msg_jam_error_s}
 	\end{align}
The maximal probability of error is then
     \begin{align}
     \maxerr(\renc,\dec)
 		&= \sup_{\jam \in \jamset(\stypes)} \max_{m \in [M]} \sum_{\vs} \msgerr(m,(\renc,\dec),\vs) \jam(\vs).
     \end{align}
The average probability of error $\avgerr(\renc,\dec)$ is defined similarly by averaging over uniform $m \in [M]$ instead of taking $\max_{m \in [M]}$.

A rate $R$ is achievable under maximum error if there is a sequence of $(n, M)$ codes $\{ (\renc_n,\dec_n) \}$ such that $\frac{1}{n} \log_2 M \ge R$ and $\maxerr((\renc_n,\dec_n)) \to 0$. The capacity $\CapwinM(\winAVC)$ is the supremum of achievable rates.
The capacity $\CapwinA(\winAVC)$ under average error is similarly defined.

\noindent \textbf{Relationship to other AVC models.} In the special case $\xwin(n) = \swin(n) = n$, the windowed AVC reduces to a (standard, windowless) oblivious AVC~\cite{DeyJLSZ:24monograph}. To state some of our results we must also introduce list codes for oblivious AVCs. An $(n,M)$ list code for an oblivious AVC $\AVC$ is a tuple $(\renc,\dec)$ where $\renc$ is defined as before and $\dec \colon \cY^n \to [M]^L$. The decoding region for message $m$ is $\mc{D}_m = \{ \vy : m \in \dec(\vy) \}$ and the errors and achievable rates are defined in the same way. We denote the maximum error capacity for list decoding with list size $L=\poly(n)$ for an AVC $\AVC$ by $\CaplistM(\AVC)$ and for list size $L = 1$ (unique decoding) by $\CapMobl(\AVC)$.
The list decoding capacity 
$ \CaplistA(\AVC) $ and unique decoding capacity $\CapAobl(\AVC)$ under average error are similarly defined. 
We note, e.g., \cite{DeyJLSZ:24monograph}, that, for  codes with probabilistic encoders, $ \CaplistM(\AVC) = \CaplistA(\AVC) $ and $\CapMobl(\AVC)= \CapAobl(\AVC)$; thus we use the notation $\Caplist(\AVC)$ and $\Capobl(\AVC)$ respectively to specify their values.

The list decoding capacity~\cite{sarwate-gastpar-2012-listdec} of a windowless AVC $\AVC$ is
\begin{align}
    \Caplist(\AVC)
    &\coloneqq \max_{P_\bfx\in\xtypes} \min_{Q_\bfs\in\stypes} I(\bfx; \bfy).\label{eqn:LD-rate} 
\end{align}
The \emph{randomized coding capacity} (with unique decoding) $\Capran$ is related: in this model Alice and Bob share a common random string of length $\Omega(\log(n))$ bits that is private from James. The randomized coding capacity for maximum and average error are equal so we use the notation $\Capran$; moreover, we also have $\Capran = \Caplist$~\cite{csiszar-narayan-it1988}~\cite[Chapter 6]{DeyJLSZ:24monograph}. This equality comes from a constuction in which Alice embeds a hash of the message into the codeword. Bob list-decodes and then discards any messages in the list that do not satisfy the embedded hash. The converse for both models follows from the converse to the channel coding theorem: if the jammer chooses to act as the DMC, he induces a channel minimizing the mutual information between the transmitter and receiver.

\begin{definition}[ECN-symmetrizability~\cite{CN:88constraints:deterministic}]
\label{def:CN-sym}
For an oblivious AVC $ \AVC = (W_{\bfy|\bfx,\bfs},\xtypes, \stypes) $, an input distribution $P_\bfx\in\xtypes $ is \emph{Ericsson--Csisz\'{a}r--Narayan (ECN) symmetrizable} if there exists a $U_{\bfs|\bfx}\in\Delta(\cS|\cX) $ with
$ \brack{U_{\bfs|\bfx} P_\bfx}_\bfs\in\stypes$ such that for all $ (x,x',y)\in\cX \times \cX \times \cY $, 
    \begin{align}
        \sum_s U_{\bfs|\bfx}(s|x') W_{\bfy|\bfx,\bfs}(y|x,s)
        &= \sum_s U_{\bfs|\bfx}(s|x) W_{\bfy|\bfx,\bfs}(y|x',s) . \notag 
    \end{align}
Let $\Delta_{\CN}(\AVC) \subset \Delta(\cX)$ be the set of all ECN-symmetrizable $P_\bfx$.
$\AVC$ is referred to as ECN-symmetrizable if and only if all $P_\bfx\in\xtypes $ are ECN-symmetrizable, i.e., $\xtypes \subseteq \Delta_{\CN}(\AVC)$.
\end{definition}

The capacity of the oblivious AVC is~\cite{CN:88constraints:deterministic}
    \begin{align}
    \Capobl(\AVC) 
    &\coloneqq \max_{P_\bfx\in\xtypes \setminus \Delta_{\CN}(\AVC)} \min_{Q_\bfs\in\stypes} I(\bfx; \bfy) , \label{eqn:CN-rate} 
    \end{align}
where the mutual information is evaluated using $P_{\bfx,\bfs,\bfy} = P_\bfx Q_\bfs W_{\bfy|\bfx,\bfs}$. If a channel is symmetrizable, the attack for James is one in which he ``spoofs'' the encoder by choosing a valid codeword $\vx'$ and passing it through the channel $U_{\bfs|\bfx'}$. The channel Bob sees is therefore a symmetric multi-access channel (MAC) and he cannot reliably distinguish between the true codeword $\vx$ and the spoofed codeword $\vx'$. The list-decoding capacity $\Caplist$ is an upper bound on the oblivious capacity $\Capobl$ and may be strictly larger if all optimizing input distributions $P_\rx$ in \eqref{eqn:LD-rate} are ECN-symmetrizable: this holds, for example, for the binary adder channel  \cite{csiszar-narayan-it1988-2}.

\section{Main Results and Motivating Example}

The main results of this work connect the list-decoding capacity $\Caplist$ of a given AVC $\AVC$ with the capacity $\Capwin$ of the corresponding windowed version $\winAVC$. In the theorems below, we prove achievability under the maximum error criteria and a matching converse under average error. This implies that 
$\CapwinA=\CapwinM$, which we denote by $\Capwin$.

\begin{theorem}\label{thm:eq}
Let $\AVC = (W_{\ry|\rx,\rs}, \xtypes, \stypes)$ be a windowless non-symmetrizable AVC. 
Let $\winAVC=(W_{\ry|\rx,\rs}, \xtypes, \stypes,\xwin,\swin)$ be a corresponding windowed AVC for which $\xwin,\swin \in [\omega(\log{n}),o(n)]$.
Then we have $\Capwin(\winAVC) = \Caplist(\AVC)$. %

\end{theorem}

\begin{remark} This work focuses on the  study of  $\xwin,\swin \in [\omega(\log{n}),o(n)]$, i.e., neither too long nor too short.
When the window constraints are very mild, i.e., $\xwin,\swin \in \Theta(n)$, then the channel starts resembling a setting with no window constraints.
On the other extreme, small window sizes (e.g., constant size) may be extremely restrictive on code design and jamming opportunities and merit an independent study. 
\end{remark}

The sufficient non-symmetrizability condition on $\AVC$ of Theorem~\ref{thm:eq} can be improved once $\xwin>\swin$. 

\begin{theorem}[$\xwin(n) > \swin(n)$]\label{thm:wxgtws}
Let $\AVC=(W_{\ry|\rx,\rs}, \xtypes, \stypes)$ be an AVC (without window constraints). 
Let $\winAVC=(W_{\ry|\rx,\rs}, \xtypes, \stypes,\xwin,\swin)$ be a corresponding windowed AVC for which $\xwin,\swin \in [\omega(\log{n}),o(n)]$ and in addition $\winratio^* \leq 1$.
Let
\begin{align}
\label{eq:gammaprime}
    \xtypes'(\winratio^*) = \left\{T_1 \mid \exists T_2\  \text{s.t.}\ \winratio^* T_1 + (1-\winratio^*) T_2 \in \xtypes\right\},
\end{align}
and let $\AVC' = (W_{\ry|\rx,\rs}, \xtypes'(\winratio^*), \stypes)$.
Then, 
if $\AVC'$ is non-symmetrizable, it holds that $\Capwin(\winAVC)=\Caplist(\AVC)$.

\end{theorem}

{We note that finding a necessary and sufficient condition under which $\winAVC$ supports non-zero rates presents some subtle technical challenges, and remains open to future work.}
Before diving into the proofs of Theorems~\ref{thm:eq} and \ref{thm:wxgtws} for general AVCs, we provide some intuition by describing our arguments specialized to bit-flipping adversaries.%

\subsection{Case study: The input-weight-constrained bit-flip AVC}
\label{sec:case}

Consider the input-weight-constrained bit-flip windowless AVC $\AVC$ in which $\cX = \cS = \cY = \{0,1\}$, the channel is given by $y = x \oplus s$ where $\oplus$ denotes binary addition, and the input $\vx$ and state $\vs$ are constrained to have Hamming weights no greater than $w$ and $p$ (both less than $1/2$) respectively, so $\xtypes = \{(1-a,a):0 \leq a \leq w\}$ and $\stypes = \{(1-b,b):0 \leq b \leq p\}$.

\noindent
\textbf{$\bullet$ ECN-symmetrizability and (non)-symmetrizability of $\AVC$:} From Definition~\ref{def:CN-sym}, $\AVC$ is ECN-symmetrizable if and only if  $w \leq p$.
Thus if $w>p$, %
communication at a positive rate $\Capobl(\AVC)>0$ is possible, \cite{CN:88constraints:deterministic}. 
If $w \leq p$, $\Capobl(\AVC)=0$. James's strategy chooses a codeword $\vx'$ uniformly at random from  Alice's codebook and sets the state sequence $\vs$ equal to $\vx'$.

\noindent
\textbf{$\bullet$ List-decoding capacity of $\AVC$ (and $\winAVC$):}
The list-decoding capacity $\Caplist$ in~\eqref{eqn:LD-rate} is $H(p \ast w) - H(p)$, where $\ast$ represents binary convolution so $p \ast w = p(1-w)+w(1-p)$. The converse comes from James simulating a binary symmetric channel with crossover $p$ and the achievability from a standard analysis of the list-decoding performance of random codes with codewords selected $\sim \mathsf{Bernoulli}(w - \varepsilon')$.
In Theorem~\ref{thm:eq}, we show that this code can be expurgated to satisfy the window constraints of $\winAVC$ as well: there are $\bigO(n)$ window constraints and the probability of a codeword violating any of these vanishes.
Thus, the list decoding capacity for $\winAVC$ is at least the list-decoding capacity of $\AVC$. \footnote{The opposite is also true, however, it is not needed to prove  Theorem~\ref{thm:eq}.}

\noindent

\noindent
\textbf{$\bullet$ Turning list-decoding for $\winAVC$ into unique decoding:}
Using standard {\em disambiguation} techniques~\cite{guruswami2003list, langberg-focs2004}, with $\bigO(\log{n})$ bits of common randomness a list code can be made uniquely decodable using a message authentication scheme.
Because of the window constraints on James and the fact that $\AVC$ is non-symmetrizable, this randomness can be communicated from Alice to Bob~\cite{ahlswede-1978}. Moreover, we can use random codes with expurgation to satisfy Alice's window-constraint.
The resulting assertion, corresponding the Theorem~\ref{thm:eq}, follows. If $w>p$ then $\CapwinM(\winAVC) \geq H(p \ast w) - H(p)$.

\noindent
\textbf{$\bullet$ Rate converse for $\winAVC$:} To show that $\CapwinA(\winAVC) \leq H(p \ast w) - H(p)$ in the window-setting, James can select $\vs$ to be i.i.d.$\sim \mathsf{Bernoulli}(p-\varepsilon')$, effectively acting as a binary symmetric channel. Since $\swin = \omega(\log(n))$, by a Chernoff bound and union bound, with high probability James will satisfy the window constraints. Thus, we can apply the converse to the channel coding theorem with high probability for a channel with inputs constrained to (relative) weight $w$.%

Theorem~\ref{thm:wxgtws} improves Theorem~\ref{thm:eq} when $\xwin>\swin$,
in the sense that a code used for $\winAVC$ in the disambiguation step can now be designed even in cases when $\AVC$ is symmetrizable.
Specifically, in this example study, the input constraint set $\xtypes'$ in \eqref{eq:gammaprime} equals $ \{(1-a,a):0 \leq a \leq w'\}$ for 
$\swin = \alpha^* \xwin$ and  $w'\triangleq  \min\{w/\alpha^*,1/2 \}$.
Let $\AVC'$ be the binary bit-flip AVC defined by $\xtypes'$ and $\stypes$; that is, codewords for $\AVC'$ are required to have Hamming weight at most $nw'$ and the jammer is constrained to flip at most $np$ bits in the transmitted codeword.
$\AVC'$ is non-symmetrizable when $w'>p$.
Notice that $w' > w$, and thus $\AVC'$ is non-symmetrizable in cases when $\AVC$ may be symmetrizable, thus the implied improvement of Theorem~\ref{thm:wxgtws} on Theorem~\ref{thm:eq}.
Below we sketch the disambiguation code design for $\winAVC$ whenever $w'>p$, i.e., whenever $\AVC'$ is not-symmetrizable.

\noindent
\textbf{$\bullet$ Turning list decoding for $\winAVC$ into unique decoding when $\xwin>\swin$:}
The main idea leading to the improved codes of Theorem~\ref{thm:wxgtws} follows the insight that windows of length $\xwin$ can convey information as long as they are non-symmetrizable in {\emph{at least one} length-$\swin$ sub-window.
More specifically, 
to share $\bigO(\log{n})$ bits of common randomness, one first considers a positive rate code for $\AVC'$.
As before, such codes can be expurgated to satisfy the requirement that any window of length $\swin$ has relative Hamming weight at most $w'$.
Refer to the resulting expurgated code as $(\renc',\dec')$.
We now modify the code $(\renc',\dec')$ by interleaving consecutive blocks of $\swin$ bits from the codewords of $(\renc',\dec')$ with the all-zero (zero cost) word of length $\left(\frac{1}{\alpha^*}-1\right)\swin$.
Codewords in the interleaved code, using the definition of $\xtypes'$, have relative weight $\leq w$.
Moreover, the interleaved code satisfies the input window-constraints of $\winAVC$ and can be used to 
share $\bigO(\log{n})$ bits for disambiguation via a decoder that only uses the symbols of the interleaved code corresponding to $(\renc',\dec')$ (i.e., the interleaved code conveys information in only a block of length $\swin$ in each window of length $\xwin$).
The resulting assertion, corresponding the Theorem~\ref{thm:wxgtws}, follows: If $w'>p$ then $\CapwinM(\winAVC) \geq  H(p \ast w) - H(p)$.
\vspace{2mm}

\section{Proof Sketch for Theorem~\ref{thm:eq}}
\label{sec:t1}

\subsection{Theorem~\ref{thm:eq}: Achievability}
We start by showing that $\CapwinM(\winAVC) \geq \Caplist(\AVC)$.
Namely, for any $R < \Caplist(\AVC)$ and any $\maxerr_\win>0$, we construct an $(n(1+o(1)),2^{Rn})$ uniquely-decodable code $(\renc_\win,\dec_\win)$ for $\winAVC$ with maximum error $\maxerr_\win$. 
At a high level, the transmission has three phases: (i) a probabilistic list-decodable codeword of length $n$, (ii) a guard window of length $\xwin$, and (iii) a hash-based disambiguation section of length $o(n)$. %

\noindent \textbf{Phase-I (Probabilistic list-decoding):} For the first phase of the transmission, we first show that for any $R < \Caplist(\AVC)$ and any $\varepsilon>0$, there exists an integer $L=\poly(1/\varepsilon)$ and an $(n,2^{Rn},L)$ list-decodable code $(\renc_\ell,\dec_\ell)$ for $\AVC$ which also satisfies the window input-constraints of $\winAVC$. 
This follows from the fact that one can design the codewords in $\renc_\ell$ in an i.i.d. fashion according to the optimizing distribution $P_\bfx$ of \eqref{eqn:LD-rate}. We assume that $P_\bfx$ is in the $\delta$-interior of $\xtypes$, i.e., any $P'_\bfx \notin \xtypes$ satisfies $\| P'_\bfx - P_\bfx \|_1 > \delta$. 
By~\cite{sarwate-gastpar-2012-listdec}, such codes attain the list-decoding capacity for a ``standard" AVC $\AVC$ (without the sliding window constraint). We then expurgate any codewords which contain a length-$\xwin$ window not satisfying the sliding window constraint. Since $\xwin \in \omega(\log(n))$, the probability of any codeword having such a window is vanishing, hence the expurgation process does not impact the rate of the designed code;
implying that  $(\renc_\ell,\dec_\ell)$ is an $(n,2^{Rn},L)$ list-decodable code for $\winAVC$ as well as $\AVC$. 

The list-codeword $\vx_\ell$ of Phase-I is now generated as follows.
For some suitably small $\varepsilon > 0$, let $r_1$ and $r_2$ be uniformly distributed over a finite field ${\mathbb F}_q$ of size $q = 2^{\varepsilon n}$.
We define the {\emph{polynomial  hash} of the encoder's message $m$ w.r.t. $r_1$ and $r_2$ as 
$h(m,r_1,r_2) = r_1 + \sum_{i=1}^{K}m_i r_2^i,$ 
where $m_i$ denotes the bits of the encoder's message $m$ decomposed into chunks of $\log(q)$ bits (with the last chunk possibly padded with $0$s as needed), $K$ is the integer $\lceil nR/\log(q)\rceil$, and all calculations are done over ${\mathbb F}_q$. We note that, even conditioned on $m$ and $h(m,r_1,r_2)$, the symbol $r_2$ is uniformly distributed over ${\mathbb F}_q$ -- this fact is useful in Phase-III.

The encoder's message for Phase-I is the concatenation $m' = m \circ h(m,r_1,r_2)$ of the message and hash. The output of the first phase encoding, $\renc_\ell(m')$, is denoted by the first phase list-codeword $\vx_\ell$. Note that $m'$ comprises of $n(R+\varepsilon)$ bits, %
hence the rate $R$ can be chosen to 
be $\Caplist - 2\varepsilon$. 
By oblivious list-decoding~\cite{DeyJLSZ:24monograph}, 
for each message $m \in \left [2^{nR}\right ]$, the  list-decoder $\dec_\ell$ outputs a list $\mc{L}$ of size $L = \poly(1/\varepsilon)$, such that with probability at least $1-\exp(-\Omega(n\varepsilon^2))$ the list $\mc{L}$ contains the correct $m \circ h(m,r_1,r_2)$.

\noindent \textbf{Phase-II (Guard window):}
For the second phase of the transmission, an arbitrary $\bar{P}_\bfx$ with rational entries is chosen from the $\delta$-interior of $\xtypes$ and a fixed deterministically-generated ``guard window'' vector $\vx_g$ of length $\xwin$ is generated such that its type equals $\bar{P}_\bfx$. In particular, without loss of generality let $\cX = \{1,\ldots,|\cX|\}$, and let $\bar{P}_\bfx$ equal $\left ( \frac{a_1}{b},\frac{a_2}{b},\ldots,\frac{a_{|\cX|}}{b},\right )$ where the $a_i$s are integers and $b = \sum a_i$. Define $K' = \lceil \xwin/b \rceil $. Then the guard-window vector $\vx_g$ comprises of $K'$ repetitions of the length $b$ vector comprising of $a_1$ occurrences of the symbol $1$, $a_2$ occurrences of the symbol $2$, etc, through to $a_{|\cX|}$ occurrences of the symbol $|\cX|$. 
By construction, the type of $\vx_g$ is in the interior of $\xtypes$ and satisfies the input window constraint. Moreover, it is not hard to verify that any sufficiently long contiguous subsequence of $\vx_g$ also has type close to $\bar{P}_\bfx$.

In Phase-II of our scheme, we concatenate $\vx_g$ to all codewords generated in Phase-I.
However, it is possible that from some list-codewords $\vx_\ell$ from the first phase, there is a length-$\xwin$ window comprising of some suffix of $\vx_\ell$ and corresponding prefix of the guard window $\vx_g$ such that this window violates the input window constraint. If so, we expurgate such list-codewords from the codebook of Phase-I. We now argue that such an expurgation (essentially) does not change the rate of the codebook from Phase-I. The argument is similar in spirit to the one made in Phase-I, however differs in details due to the different generation processes for $\vx_\ell$ and $\vx_g$.

Let $P_\bfx$ be the type used to construct $\vx_\ell$ in Phase-I.
Note that both $P_\bfx$ and $\bar{P}_\bfx$ are in the $\delta$-interior of $\xtypes$. This implies that for any length-$\xwin$ window whose overlap with $\vx_g$ is of length at least $(1-\varepsilon)\xwin$ (and therefore whose overlap with some $\vx_\ell$ is of length at most $\varepsilon \xwin$), the type of the subcodeword within this window is also within $\xtypes$. Similarly, for any length-$\xwin$ window whose overlap with $\vx_g$ is of length at most $\varepsilon\xwin$, the type of the subcodeword within this window is also within $\xtypes$. 
Thus, in the expurgation process in this phase, we therefore only have to consider windows such that the suffix from some $\vx_\ell$ is of length at least $\varepsilon \xwin$ and at most $(1-\varepsilon)\xwin$. But since each list codeword is chosen i.i.d., by standard concentration inequalities with probability $1-\exp(-\Omega(\xwin\varepsilon^2))$ the type of this suffix is also within $\xtypes$. In addition, as  argued above, the type of any long prefix of $\vx_g$ is close to $\bar{P}_\bfx$, and hence also within $\xtypes$. Therefore, since $\xtypes$ is a convex set, the type of such an overlapping window must also be within $\xtypes$.

Taking a union bound over all at most $\xwin$ windows that have overlap between $\vx_\ell$ and $\vx_g$ we note that the expected fraction of list-codewords that would need to be expurgated in this phase is at most $\xwin\exp(-\Omega(\xwin\varepsilon^2))$. Since $\xwin \in \omega(\log(n))$, this decays faster than any polynomial in $n$, so certainly faster than say $1/n$. Since each codeword in the list codebook is generated independently, therefore by the Chernoff bound, with probability superexponentially close to one no more than a $2/n$ fraction of list-codewords have to be expurgated.

\noindent \textbf{Phase-III (Hash transmission):} In the third phase we now use a standard oblivious communication scheme to communicate to the decoder the two random keys $r_1$ and $r_2$ used in the first phase. This requires transmitting a message of $2\varepsilon n$ bits over a non-symmetrizable channel with capacity $C=\Capobl$. To do so we use as a black box the codes from~\cite{csiszar-narayan-it1988-2}, with blocklength $2n\varepsilon/(C - \varepsilon)$, with codewords drawn i.i.d. according to the maximizing non-symmetrizable distribution given in~\eqref{eqn:CN-rate}. As in  Phase-I and Phase-II discussed above, we also expurgate codewords from this oblivious communication scheme so as to ensure no codeword violates its window constraint. These arguments are identical to the ones before and omitted.

The decoder, on obtaining $r_1$ and $r_2$ then uses them to disambiguate the list from Phase-I down to a unique message. 
By the Schwartz-Zippel lemma, the probability that a message other than the true message also satisfies the hash in Phase-I is exponentially small in $n$.

\subsection{Theorem~\ref{thm:eq}: Converse}
We now address the proof for the assertion $\CapwinA(\winAVC) \leq \Caplist(\AVC)$.
The proof uses the fact \cite{sarwate-gastpar-2012-listdec} that, for $R >\Caplist(\AVC)$, any $(n,2^{Rn},L)$ list-code 
$(\renc,\dec)$ for $\AVC$ must have average error $\avgerr=\Omega(1)$ even under a jamming strategy $\jam$ for which for any $\vs=(s_1,\dots,s_n)$,
$\jam(\vs) = \prod_{i=1}^n P_{\bfs}(s_i)$
for a suitable distribution $P_{\bfs}$ over $\cS$.
Namely, the jamming strategy $\jam$ chooses a state vector $\vs$ letter-by-letter in an i.i.d. manner via $P_{\bfs}$.
One can now use $\jam(\vs)$ on $\winAVC$ as well to cause $\avgerr_\win=\Omega(1)$. 
Roughly speaking, as the choice of $\vs$ is done in an i.i.d. manner, using standard concentration bounds combined with a union bound of all state-windows, the resulting $\vs$, for $\swin=\omega(\log{n})$, satisfies all state-window constraints. 
This implies that this jamming strategy $\jam$ when used on $\winAVC$  causes a constant average decoding-error on a given rate $R$ encoder (for list-decoding and thus also for unique-decoding).

\section{Proof sketch for Theorem~\ref{thm:wxgtws}}
\label{sec:t2}

We now sketch the major modifications needed in the proof of Theorem~\ref{thm:eq} to obtain Theorem~\ref{thm:wxgtws}.
First note that the converse proof of Theorem~\ref{thm:eq} does not require $\AVC$ to be non-symmetrizable and thus applies to  Theorem~\ref{thm:eq} as well.
The achievability scheme here has an overall similar 3-phase structure as in the proof of Theorem~\ref{thm:eq}, with Phase-I being exactly the same. We describe Phase-II (guard window) and Phase-III (hash transmission) below.

We consider a large enough $n$ such that $\winratio = \swin/\xwin$ is close enough to $\winratio^*$ such that $(W_{\ry|\rx,\rs}, \xtypes'(\winratio), \stypes)$ is also non-symmetrizable.
For Phase-II and Phase-III, a non-symmetrizable distribution $T_1$ is chosen from the interior of $\xtypes'(\winratio)$. A corresponding $T_2$ is also chosen, such that $\alpha T_1 + (1-\alpha)T_2$ is in the interior of $\Gamma$.  We assume that $\alpha T_1$ and $(1-\alpha)T_2$ have rational components with denominators dividing $\xwin$. We choose an integer $\runstep=\lambda \alpha \xwin$ for some sufficiently small constant $\lambda>0$. 

In what follows, we construct the remaining portions of the code corresponding to Phase-II and III. In our design, both portions consist of an integer number of disjoint windows of length $\xwin$. Moreover, each disjoint window of length $\xwin$ has some positions corresponding to words of type $T_1$; and other positions corresponding to words of type $T_2$. The former set of positions is denoted by $S_1$ and the latter set  by $S_2$. As the positions of $S_1$ and $S_2$ may differ for different windows, our notion may possibly have an additional subscript to indicate the window index. 
We will call the locations in $S_1$ and $S_2$ as respectively type-1 and type-2 locations. In each disjoint window of length $\xwin$, the fractions of locations in $S_1$ and $S_2$ are respectively $\alpha$ and $1-\alpha$, so that the total type in the window of length $\xwin$ is $\alpha T_1+ (1-\alpha)T_2 \in \Gamma$. Next we will describe in detail how the $S_1$ and $S_2$ sets are chosen in each disjoint window in each phase.

Phase-II has $1+\lceil\frac{1}{\lambda}\rceil$ disjoint windows of length $\xwin$ each, hence the total length is $\xwin (1+\lceil\frac{1}{\lambda}\rceil)$. The windows are indexed by $i=0,1,\ldots,\lceil\frac{1}{\lambda}\rceil$. For the $i$'th window, the $\xwin$ locations are partitioned into two disjoint subsets $S_{1,i},S_{2,i}$. In the $i$'th window, the first $L_i=\min\{i\runstep,\alpha\xwin\}$ positions are set to be in $S_{1,i}$ and the next $\lceil\frac{(1-\alpha)L_i}{\alpha}\rceil$ positions are in $S_{2,i}$. This ensures that the first $L_i+\lceil\frac{(1-\alpha)L_i}{\alpha}\rceil$ locations are distributed in the ratio $\alpha:(1-\alpha)$ between the subsets $S_{1,i}$ and $S_{2,i}$. For the remaining locations, we allocate sequentially and put the $j$'th location in $S_{2,i}$ if and only if the fraction of preceding positions (positions $1$ to $j-1$) in $S_{2,i}$ is less than $1-\alpha$, otherwise we put $j$ in $S_{1,i}$. At the end of this allocation, we have $|S_{1,i}|:|S_{2,i}|=\alpha:(1-\alpha)$. Note that, in the last window in Phase-II, i.e. for $i=\lceil\frac{1}{\lambda}\rceil$, the first $\alpha\xwin$ positions are in $S_{1,i}$ and the rest of the positions are in $S_{2,i}$. It can be checked that in any sliding window of length $\xwin$ in Phase-II, the fraction of type-1 locations is between $\alpha$ and $\alpha(1 + \lambda)$. 

In Phase-III, each disjoint window is divided between $S_{1}$ and $S_{2}$ in a similar manner as in the last window of Phase-II. That is, in each of the disjoint windows in the third phase, the first $\alpha\xwin$ locations are in $S_{1}$ and the rest of the locations are in $S_{2}$. Hence in any sliding window of length $\xwin$ overlapping with Phase-III, the fraction of type-1 locations is exactly $\alpha$.

We now describe how transmitted symbols are chosen for type-1 and type-2 positions in each phase. For all windows in Phase-II and Phase-III, all type-2 positions are considered as a long sequence. These positions are allotted symbols from $\cX$ deterministically according to $T_2$ in a similar manner as in the guard window in the proof of Theorem~\ref{thm:eq}. These fixed symbols are transmitted in those locations. For all the type-1 positions in Phase-II also, symbols are allotted from $\cX$ deterministically according to $T_1$ in a similar manner.   Since in all the windows in Phase-II, the transmission is deterministic, the type in any sliding window deviates from $\alpha T_1 +(1-\alpha) T_2$ by at most $\lambda\alpha$. Since $\alpha T_1 +(1-\alpha) T_2$ is in the interior of $\Gamma$, the input window constraint is satisfied for a small enough choice of $\lambda$. 

We are left to determine the transmitted symbols in all the type-1 positions in Phase-III.
Recall from the proof of Theorem~\ref{thm:eq}, the use of a code generated at random (and then expurgated) for the transmission of the random keys $(r_1,r_2)$.
A similarly positive rate code is used here as well, however with codewords with entries distributed according to $T_1$.
Specifically, disambiguation will use the type-1 positions in Phase-III  that hold a codeword in the above code corresponding to $(r_1,r_2)$.
Following the proof of Theorem~\ref{thm:eq}, standard concentration and union bounds over sliding windows are used to guarantee that, after an appropriate expurgation process, the resulting code, on Phases II-III, satisfy the sliding input window constraints.

\section*{Acknowledgements}

B.~K.~Dey was supported in part by the Bharti Centre for Communication in IIT Bombay.
The work of M. Langberg was supported in part by NSF grant CCF-2245204. The work of S. Jaggi was supported by the Engineering and Physical Sciences Research Council [Grant Ref: EP/Y028732/1]

\clearpage

\bibliographystyle{IEEEtran}
\bibliography{ref}

\end{document}